\documentclass[12pt]{article}

\usepackage[margin=1in]{geometry}
\usepackage{cite}
\usepackage{amsmath,amssymb,amsfonts}
\usepackage{algorithmic}
\usepackage{graphicx}
\usepackage{textcomp}
\usepackage{xcolor}
\usepackage{multicol}
\usepackage{multirow}
\usepackage{listings}
\usepackage{booktabs}
\usepackage{caption}
\usepackage{bbm}
\usepackage{hyperref}

\title{Reliable Curation of EHR Dataset via Large Language Models under Environmental Constraints}

\author{
	Raymond M. Xiong,
	Panyu Chen,
	Tianze Dong,
	Jian Lu,\\
	Louis Hu, Nathan Yu, Benjamin Goldstein, Danyang Zhuo,\\
	and 
	Anru R. Zhang\\
	Duke University
}

\date{} 

\begin{document}
	\maketitle
	
\begin{abstract}
	Electronic health records (EHRs) are central to modern healthcare delivery and research; yet, many researchers lack the database expertise necessary to write complex SQL queries or generate effective visualizations, limiting efficient data use and scientific discovery. To address this barrier, we introduce CELEC, a large language model (LLM)-powered framework for automated EHR data extraction and analytics. CELEC translates natural language queries into SQL using a prompting strategy that integrates schema information, few-shot demonstrations, and chain-of-thought reasoning, which together improve accuracy and robustness. CELEC also adheres to strict privacy protocols: the LLM accesses only database metadata (e.g., table and column names), while all query execution occurs securely within the institutional environment, ensuring that no patient-level data is ever transmitted to or shared with the LLM. On a subset of the EHRSQL benchmark, CELEC achieves execution accuracy comparable to prior systems while maintaining low latency, cost efficiency, and strict privacy by exposing only database metadata to the LLM. Ablation studies confirm that each component of the SQL generation pipeline, particularly the few-shot demonstrations, plays a critical role in performance. By lowering technical barriers and enabling medical researchers to query EHR databases directly, CELEC streamlines research workflows and accelerates biomedical discovery.
\end{abstract}
	
	
\begin{sloppypar}
\section{Introduction}
\label{sec:intro}

Electronic health records (EHRs) have become a central component of modern healthcare. Since 2009, EHR adoption has increased tenfold, and by 2020, more than 90\% of primary care providers worldwide reported daily use \cite{barker2024evolution, kerr2024features}. When effectively implemented, EHRs improve the accuracy and continuity of patient care while enabling large-scale, cost-efficient data analysis that supports both clinical research and population health management \cite{casey_electronic_2016, kruse_use_2018, chishtie_use_2023}. Unlocking this potential requires methods that make EHR data accessible and interpretable for research, advancing both scientific discovery and healthcare systems.

In practice, however, EHR systems remain challenging for researchers to use. Extracting meaningful data often requires navigating fragmented interfaces and applying technical skills such as writing structured queries or custom analytics, which most researchers lack formal training in \cite{tsai2020effects, kruse_use_2018}. These barriers slow scientific progress, limit exploratory analysis, and necessitate reliance on database specialists, leaving valuable research opportunities underutilized. Tools that allow researchers to access EHR data and conduct basic analytics without programming expertise could therefore lower entry barriers and accelerate biomedical discovery \cite{honeyford2022challenges}.

Prior work has attempted to bridge this gap using text-to-SQL methods. While these approaches have shown strong performance on EHR question-answering benchmarks \cite{jo-etal-2024-lg, gundabathula-kolar-2024-promptmind, kim-etal-2024-probgate, attrach2025conversational}, it is often unclear whether the underlying database systems are exposed to cloud-based LLMs, which can be a significant risk for sensitive medical datasets, where data leakage might compromise patient privacy.

To address the security concern, we propose \textbf{CELEC (\underline{C}uration of \underline{E}HR via \underline{L}LMs under \underline{E}nvironmental \underline{C}onstraints)}, a system that restricts LLM access to schema-level metadata only and executes all SQL queries locally, ensuring that no patient-level data is transmitted or shared inadvertently with external services. As a framework for automated EHR data extraction and analytics, CELEC translates natural-language (NL) questions into SQL using a prompting strategy that combines schema information, few-shot demonstrations, and chain-of-thought reasoning to generate accurate and robust SQL outputs.

Once data are retrieved, CELEC can also generate simple visualizations directly from the extracted dataframes, providing immediate exploratory insights. Designed with real-world constraints in mind, CELEC maintains low latency and cost efficiency. CELEC also adheres to strict privacy protocols: the LLM interacts only with database metadata (e.g., table and column names), while all query execution occurs securely within the institutional environment. At no point is patient-level data transmitted to or shared with the LLM. By lowering technical barriers, CELEC enables researchers to query EHR databases and perform preliminary analyses independently, thereby streamlining workflows and accelerating scientific discovery.

\section{Related Work}
\label{sec:lit-review}

\subsection{EHR system usability \& data accessibility}

Despite near-universal adoption of EHRs \cite{barker2024evolution, shen_twenty-five_2025}, effective use remains limited by fragmented information, disrupted workflows, and high cognitive load, which are commonly linked to errors, inefficiency, and burnout among both clinicians \cite{asgari_impact_2024, olakotan_usability_2025}. Prior research has shown that extensions such as custom dashboards and enhanced visualization of clinical trends can improve usability \cite{mazur_association_2019, pollack_association_2020}. More recently, automated visualization approaches have been explored in the medical domain \cite{zhang2025visualizingelectronicmedicalrecords}; however, these efforts often focused on small-scale applications and electronic medical records rather than large-scale, multi-table EHR systems. 

Beyond usability in care delivery, secondary use of EHR data is essential for research, quality improvement, and population health management. Nevertheless, accessibility remains a major barrier, as healthcare professionals often lack the training required to manipulate complex database systems \cite{kruse_use_2018, tsai2020effects}. To reduce this barrier, several interactive EHR query tools have been proposed. For example, i2b2\cite{murphy2010i2b2} provides a drag-and-drop cohort builder in which users browse a hierarchical ontology of clinical concepts (e.g., diagnoses, laboratory tests, medications, and procedures) and place them into panels such as ``Inclusion Criteria'' and ``Exclusion Criteria'' to construct a query. Subsequent systems, such as LEAF\cite{dobbins2020leaf} and OMOP/ATLAS\cite{hripcsak2015ohdsi}, follow a similar paradigm, augmenting it with modern web interfaces and richer configuration options to make the experience more intuitive. While these tools eliminate the need to write SQL explicitly, they still require users to think in terms of database structure and to formulate queries using controlled vocabularies and formal logic. As a result, effective use often demands substantial familiarity with the underlying data model and query logic, which limits accessibility for users with minimal exposure to databases and programming and constrains rapid exploratory analysis. 

In contrast, the goal of CELEC is to advance accessibility to the next level. CELEC removes the need for users to understand query logic, database structure, or any programming language. Instead, users simply pose a natural-language question, and the system automatically retrieves relevant data, performs the necessary analysis, and produces both tabular results and visual summaries. By unifying natural language querying with flexible analytics and visualization within a single pipeline, CELEC provides a deployment-ready framework that directly supports research-driven use cases. This question-driven interaction paradigm offers a substantially more intuitive experience than schema-aware query builders and fundamentally distinguishes CELEC from existing interactive EHR tools.

\subsection{Medical text-to-SQL}

General-purpose text-to-SQL models have achieved strong results on benchmarks such as Spider \cite{yu2019spider} and WikiSQL \cite{zhong2017seq2sql} \cite{gao2023texttosqlempoweredlargelanguage, pourreza2023dinsqldecomposedincontextlearning, li2024petsqlpromptenhancedtworoundrefinement}. However, performance on these datasets does not guarantee success in clinical contexts, where data complexity and privacy requirements differ substantially from business applications. Early medical-domain systems relied on templates for text-to-SQL translation, which limited their ability to address complex real-world questions \cite{wang2020texttosqlgenerationquestionanswering}. The EHRSQL shared task \cite{lee2023ehrsqlpracticaltexttosqlbenchmark, lee-etal-2024-overview} advanced the field by collecting natural language queries from hospital staff and aligning them to databases such as MIMIC-III, eICU, and MIMIC-IV demo. Leading teams on the benchmark explored methods including schema-aware models \cite{jo-etal-2024-lg}, ensemble prompting \cite{gundabathula-kolar-2024-promptmind}, and probability-based SQL verification \cite{kim-etal-2024-probgate}. CELEC employs a streamlined two-call design that balances accuracy and low latency while strictly adhering to privacy protocols. Relevant quantitative evaluations are presented in Section~\ref{sec:eval}.

\subsection{LLMs in Healthcare}

Most LLM applications in healthcare have centered on clinicians for decision support, summarization, and conversational interfaces \cite{oniani2024enhancinglargelanguagemodels, rajashekar2024humanalgo}. However, comparatively little attention has been given to applications that assist researchers with structured data analytics. Recent work has explored connecting EHR databases with Claude Desktop for LLM-powered analytics \cite{attrach2025conversational}. In contrast, CELEC introduces a privacy-conscious design that restricts LLM inputs to schema-level metadata, thereby minimizing exposure of raw patient data and reducing the attack surface for leakage.

\begin{figure*}[htbp]
	\centering
	\vspace{-10pt}
	\includegraphics[width=\textwidth]{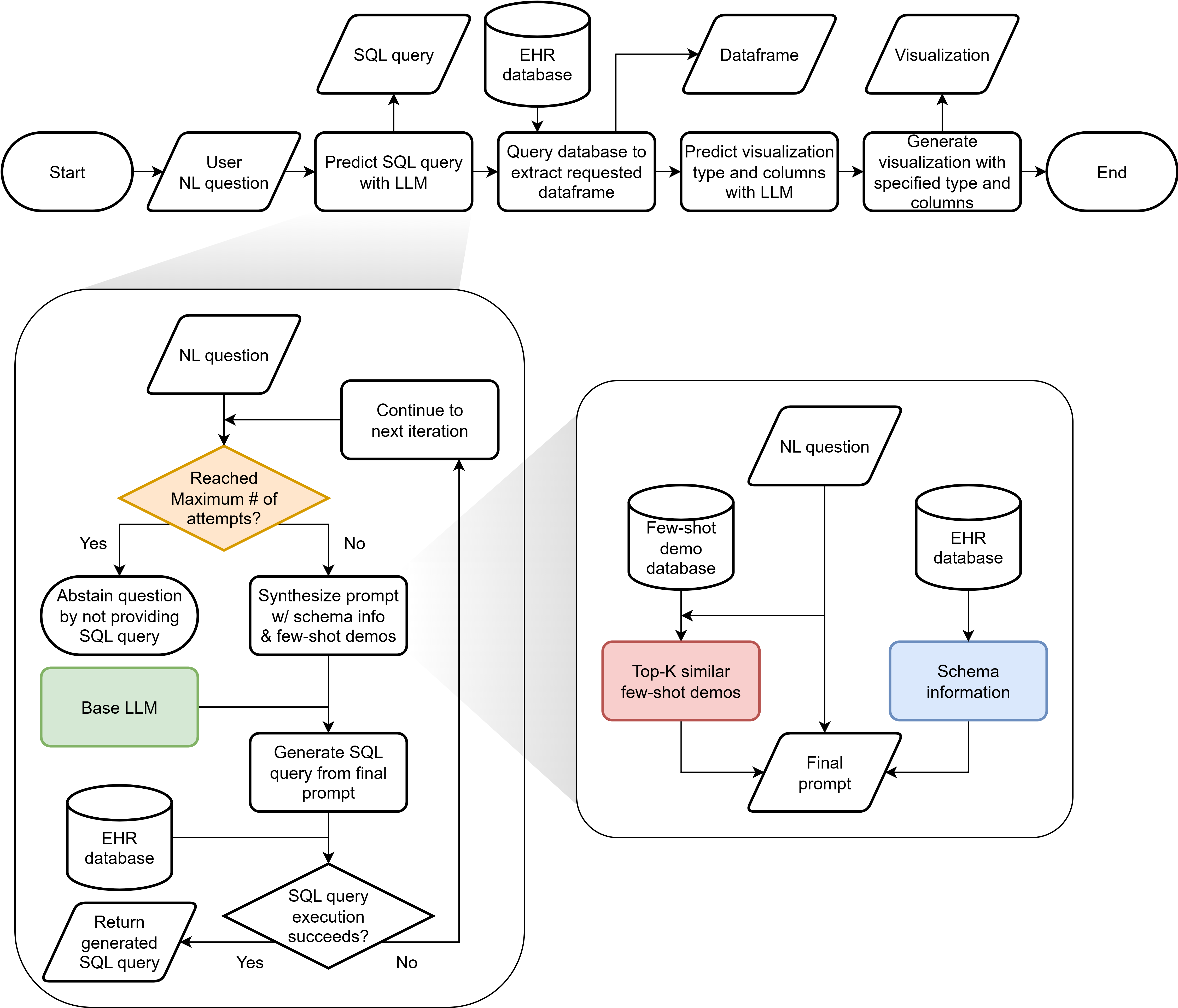}
	\caption{Overview of the CELEC system design.}
	\label{fig:system}
	\vspace{-15pt}
\end{figure*}

\section{Methods}
\label{sec:methods}

We designed the CELEC system\footnote{The source code and data will be released through GitHub.} as an end-to-end application (see Figure \ref{fig:system}). The system will connect to an EHR database. In our implementation, we connect to DuckDB versions of MIMIC-III \cite{mimic_iii} and MIMIC-IV \cite{johnson10mimic} databases, which are large EHR databases comprising de-identified health-related data, including demographics, measurements, laboratory test results, procedures, and medications. Users can enter a natural-language (NL) question as input to specify the objective of their data extraction or analysis. Then, we use an LLM to translate the NL question into an SQL query that extracts a succinct yet necessary set of data to answer it. We use prompt engineering to enhance the quality of generated SQL code (Section~\ref{met:sql-gen}). Next, we connect to the EHR database and execute the generated SQL to retrieve the corresponding data. After the requested dataframe is retrieved, we use an LLM to create a visualization that effectively answers the NL question or enables users to better explore their data (Section~\ref{met:viz-gen}).

\subsection{SQL generation}
\label{met:sql-gen}

We use the {\tt o3-2025-04-16} model to generate SQL queries from NL input. To improve reliability and alignment with the target EHR database, our prompt design incorporates three key elements: (1) \textbf{schema information}, (2) \textbf{few-shot demonstrations}, and (3) \textbf{chain-of-thought (CoT)} reasoning. The complete prompt template is provided in Appendix~\ref{apd:prompt-template-sql}. Importantly, only schema-level metadata (table names, column names, and column types) are passed to the LLM; no patient-level data are ever exposed to the LLM.

\textbf{Schema information.} For each database, schema metadata are embedded in the prompt to inform the LLM of available tables and attributes, which reduces hallucination and prevents fabrication of nonexistent aliases. Each table is presented along with its columns, declared data types, and relevant constraints (e.g., primary keys). More details may be found in Appendix~\ref{apd:prompt-template-sql}.  

\textbf{Few-shot demonstrations.} We include in-context few-shot demonstrations of NL questions paired with their corresponding SQL queries. Demonstrations come from two sources: (a) medical literature using MIMIC datasets, where questions were adapted from published cohort criteria and manually verified (105 demos; see Appendix \ref{apd:medical-lit-demo}), and (b) the EHRSQL benchmark \cite{lee-etal-2024-overview}, from which we preprocessed the training and validation sets into 4,761 high-quality pairs (see Appendix \ref{apd:test-set-sampling}). These sources are combined into a single demo database; during inference, CELEC selects the top-$k$ most similar demos ($k=2$) based on cosine similarity between the input question's embedding and the embeddings of the demo questions. Embeddings are computed using {\tt all-MiniLM-L6-v2} and indexed with ChromaDB for efficient retrieval.  

\textbf{Chain-of-thought (CoT).} To guide complex query construction, we augment the prompt with an intermediate reasoning step where the LLM first identifies potentially relevant tables before generating the final SQL. This step enhances alignment between the input question and the selected schema elements, thereby reducing unnecessary joins and extraneous columns. Crucially, the few-shot demonstrations are also formatted to include this intermediate reasoning. By showing demos that explicitly map NL questions to table-selection steps and then to final SQL, we encourage the model to follow the same structure. This not only enhances interpretability but also stabilizes SQL generation in complex queries.

\textbf{Error handling.} After SQL generation, the query is executed locally on the target database. If execution fails (typically due to hallucinations of aliases or function names), the error message is appended to the prompt, and the LLM retries the generation. By default, up to two retries are allowed, which balances robustness with latency (see ablation results in Section~\ref{sec:ablation}). Error messages are processed only by the LLM during retries and are not exposed to users.  

\subsection{Visualization generation}
\label{met:viz-gen}

As many research and clinical workflows benefit from visual summaries that facilitate pattern interpretations, CELEC also includes an LLM-powered visualization module. After the SQL query is executed and a dataframe retrieved, we provide the LLM with the input NL question and the column metadata from the resulting table. The model predicts (1) an appropriate visualization type (e.g., histogram, bar chart, line plot, scatterplot) and (2) the aesthetic mappings, such as which variables should be plotted on the horizontal and vertical axes. 

The prompt includes both instructions and simple input-output examples (see Appendix~\ref{apd:prompt-template-viz}), encouraging the model to output a structured specification. Importantly, as in SQL generation, only schema-level metadata is exposed to the LLM; the underlying patient-level data remains inaccessible. Once the visualization specification is returned, CELEC renders the chart using a set of hard-coded TextScript functions. This design ensures consistency and robustness across visualizations, while protecting privacy and minimizing LLM-induced errors. By integrating visualization generation, CELEC enables users to move beyond raw tabular results and obtain exploratory insights directly, thereby lowering the barriers for both researchers and clinicians.

\subsection{System integration}
\label{met:integration}

While Sections~\ref{met:sql-gen} and \ref{met:viz-gen} describe SQL and visualization generation individually, their integration is key to CELEC's usability in real-world research settings. The modules are orchestrated in a unified pipeline: SQL queries are generated and executed locally, and results can either be returned directly or forwarded to the visualization module for immediate exploratory analysis.

This integration offers system-level benefits that neither module alone captures. First, the streamlined design, which relies on only two LLM calls per query, reduces latency compared to ensemble or verification-heavy approaches. Second, the privacy-conscious architecture ensures that at no stage do LLMs access patient-level data; only schema metadata and aggregate specifications are processed. Finally, presenting SQL and visualization under a single framework reduces the technical barrier for researchers and clinicians alike, supporting both reproducible research and practical deployment.

\section{Evaluation}
\label{sec:eval}

We evaluate CELEC on the EHRSQL-2024 benchmark \cite{lee-etal-2024-overview}, which contains more than 7,000 pairs of NL questions and gold SQL queries against a modified version of the MIMIC-IV demo database. The benchmark is designed to reflect real-world clinical needs, as its queries are typically complex, involving temporal conditions, group operations, and cross-table joins. 

\subsection{Dataset adaptation}
To align the benchmark with CELEC, we filtered and modified the provided train, validation, and test splits. In particular, we removed unanswerable questions where the gold SQL did not execute successfully on the MIMIC-IV demo database (see Appendix~\ref{apd:test-set-sampling} for full details). After adaptation, we used all training and validation points as few-shot candidates for SQL prompting, 10\% of the new test set (78 queries) as validation data for hyperparameter tuning, and the remaining 90\% (707 queries) for evaluation. By ensuring that all test queries are executable, this adaptation yields a more reliable measure of model capability. We note that the deviation from the official leaderboard split precludes direct performance comparisons between CELEC and published systems; we present the leaderboard numbers as a contextual reference for understanding CELEC's performance. 

\subsection{Evaluation methods}
We evaluate system performance using the RS(0) score, the official metric of the EHRSQL benchmark. RS(0) extends execution accuracy (EX) by rewarding abstention on unanswerable questions. Formally, if $X$ is the set of all questions and $X_{ans} \subseteq X$ the subset of answerable ones, then

\[
\begin{aligned}
	RS(0) & = \frac{1}{|X|}\sum_{x \in X}\mathbbm{1}\left[(x \in X_{ans} \land \big.R_{gen}(x)=R_{gt}(x))\right.\\
	& \left. \lor (x \notin X_{ans} \land g(x)=0)\right].
\end{aligned}
\]

Since our adapted test set excludes unanswerable questions ($X_{ans}=X$), RS(0) reduces to execution accuracy in our evaluation, but we report it as RS(0) for consistency with leaderboard scores.

\begin{table}[t]
	\centering
	\caption{Comparison of CELEC with teams from the EHRSQL-2024 leaderboard.}
	\label{tab:ehrsql-full}
	\begin{tabular}{lcc}
		\toprule
		\textbf{System / Team} & & \textbf{RS(0) (\%)}\\
		\midrule
		\textbf{CELEC (Ours)} & & \textbf{81.05}\\
		\midrule
		LG AI Research \& KAIST & \cite{jo-etal-2024-lg} & 88.17\\
		PromptMind & \cite{gundabathula-kolar-2024-promptmind} & 82.60\\
		ProbGate & \cite{kim-etal-2024-probgate} & 81.92\\
		GPT-4o & \cite{xu_medagentgym_2025} & 76.45\\
		KU-DMIS & \cite{park-etal-2024-ku}   & 72.07\\
		AIRI NLP & \cite{somov-etal-2024-airi} & 68.89\\
		MedCopilot & \cite{xu_medagentgym_2025} & 68.78\\
		LTRC-IIITH & \cite{thomas-etal-2024-ltrc} & 66.84\\
		Text2SQL-Flow & \cite{cai_text2sql-flow_2025} & 58.70\\
		Saama Technologies & \cite{jabir-etal-2024-saama} & 53.21\\
		\bottomrule
	\end{tabular}
	\vspace{-10pt}
\end{table}

\begin{table*}[htbp]
	\centering
	\vspace{-10pt}
	\caption{RS(0) score of CELEC under different parameter settings. The first row represents our default settings.} 
	\label{tab:ablation}
	\resizebox{0.85\textwidth}{!}{%
		\begin{tabular}{cccc|c}
			\toprule
			\textbf{Base LLM} & 
			\textbf{Schema info? (Y/N)} & 
			\textbf{\# of few-shot demos} & 
			\textbf{Max \# of attempts} & 
			\textbf{RS(0) (\%)} \\
			\midrule
			\textbf{o3}       & \textbf{Y} & \textbf{2} & \textbf{2} & \textbf{81.05} \\
			
			\midrule
			
			\textbf{o4-mini}  & Y & 2 & 2 & \textbf{73.41} \\
			\textbf{gpt-4.1}  & Y & 2 & 2 & \textbf{77.93} \\
			
			\midrule
			
			o3       & \textbf{N} & 2 & 2 & \textbf{77.93} \\
			
			\midrule
			
			o3       & Y & \textbf{1} & 2 & \textbf{73.97} \\
			o3        & Y & \textbf{0} & 2 & \textbf{50.21} \\
			
			\midrule
			
			o3        & Y & 2 & \textbf{1} & \textbf{79.21} \\
			\bottomrule  
		\end{tabular}%
	}
	\vspace{-10pt}
\end{table*}

\subsection{Results \& Discussion}
Table~\ref{tab:ehrsql-full} compares CELEC with leading systems from the EHRSQL-2024 leaderboard. On our adapted test set, CELEC achieves 81.05\% RS(0) accuracy. This result highlights CELEC's ability to achieve accuracy comparable to existing systems while maintaining strong privacy guarantees. Ablation studies in Section~\ref{sec:ablation} further confirm that few-shot prompting and retry mechanisms are critical to this performance. 

In terms of latency, CELEC processes one NL query in an average of 6.02 to 6.11 seconds across the test set. As for API costs, CELEC incurs an average cost of \$0.0152 per question across the test set.  While other systems have not disclosed inference times and associated costs, CELEC's streamlined two-call design demonstrates practical feasibility for interactive use and large-scale deployment.

Finally, to safeguard privacy, we verified automatically that all generated SQL queries only reference schema-level columns and tables. No queries attempted to access fabricated or patient-level identifiers, consistent with CELEC's metadata-only exposure design.

\section{Ablation Study}
\label{sec:ablation}

We conducted ablation studies to assess the contribution of individual system components. In each experiment, we modified one component while keeping all others fixed, and report the RS(0) score on the adapted EHRSQL test set (Table~\ref{tab:ablation}). Results consistently show that each component makes a meaningful contribution to performance, with in-context learning yielding the largest gains.

\textbf{Base LLM.}  
The choice of the base model had a great impact on accuracy. Our default model, {\tt o3-2025-04-16}, achieved the highest score of 81.05\%. Replacing it with the smaller reasoning model {\tt o4-mini-2025-04-16} reduced accuracy to 73.41\%, whereas using {\tt gpt-4.1-2025-04-14}, a non-reasoning model, yielded 77.93\%. This suggests that both model scale and reasoning optimization are essential for handling the complex queries in EHRSQL.

\textbf{Schema information.}  
Providing schema metadata to the LLM significantly improved reliability. Without schema information, performance dropped from 81.05\% to 77.93\%, with frequent hallucinations of nonexistent columns or aliases. This confirms that grounding query generation in explicit schema details is critical to reducing structural SQL errors.

\textbf{Few-shot demonstrations.}  
In-context demonstrations had the most significant impact of all components. Without demonstrations (zero-shot), CELEC achieved only 50.21\%. Adding a single demonstration increased performance to 73.97\%, and using two demonstrations, which is our default, further improved it to 81.05\%. This trend underscores the crucial role of in-context learning in aligning the model with the structure and semantics of medical databases, with diminishing returns evident beyond two examples.

\textbf{Maximum attempts.}  
Allowing the model to retry once after a failed SQL execution improved accuracy from 79.21\% (single attempt) to 81.05\% (two attempts). Error analysis indicates that many first-pass failures stemmed from superficial issues, such as hallucinations of aliases or function names. Providing the database error message during a second attempt often corrected these, yielding modest but consistent gains with minimal latency overhead.

Overall, these results demonstrate that each design choice contributes to CELEC's performance, with few-shot demonstrations being the most critical factor. The combination of schema grounding, a reasoning-capable base model, and a lightweight retry mechanism together supports robust SQL generation in complex EHR environments.

\section{Error Mode Analysis}
\label{sec:error-mode}

\begin{figure}[!htbp]
	\centering
	\vspace{-10pt}
	\includegraphics[width=\linewidth]{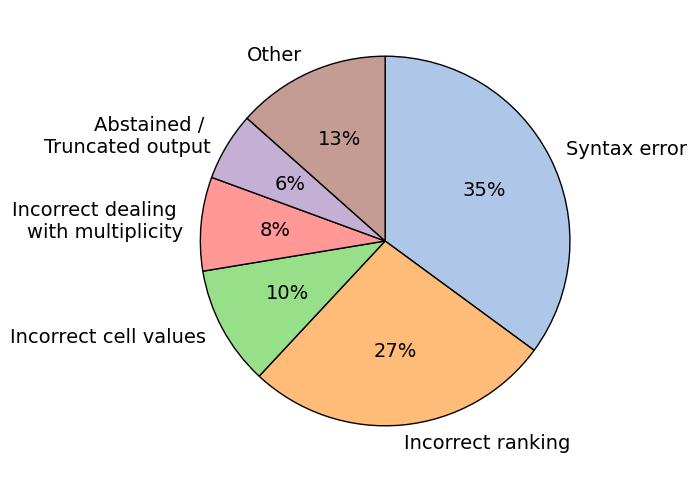}
	\caption{Statistics of failure modes under default settings.}
	\vspace{-10pt}
	\label{fig:error-mode}
\end{figure}

To better understand the patterns underlying failed queries, we conducted an error mode analysis. We manually inspected cases where the result table produced by the predicted SQL did not match the ground-truth output and categorized the observed failures. Several recurring error modes emerged, offering insights into the model's limitations and suggesting concrete directions for future improvement (Figure \ref{fig:error-mode}).

\textbf{Syntax error.}
Syntax errors were the most frequent error mode we observed, accounting for 35\% of failures. In these cases, the predicted queries commonly included either undefined functions (e.g., ``\texttt{date}'', which is not a DuckDB function) or invalid function argument types (e.g., for the \texttt{date\_diff} function). A likely explanation is confusion with functions from more traditional SQL dialects, in which these queries would be valid. As these newer dialects gradually gain popularity in various applications for their expressive power and performance benefits, future work in general-domain Text-to-SQL could consider more explicit modeling of dialectal variation, potentially by developing new benchmarks.

\textbf{Incorrect ranking.}
Incorrect ranking was the second most common error type, accounting for 27\% of all failures. These errors systematically occurred for questions that requested a fixed number of ranked results, typically phrased as ``what are the top K items...''. In the ground-truth SQL, such queries often rely on the \texttt{DENSE\_RANK} function and filter results by ranks no greater than K, which allows the output to include more than K rows when ties occur at the cutoff. In contrast, the predicted SQL queries frequently use \texttt{ORDER\_BY} in conjunction with \texttt{LIMIT}, which enforces an output of exactly K rows and thus fails to account for ties. This discrepancy suggests a limitation of the system in handling multiple valid interpretations of NL ranking questions. Future work could explore uncertainty-aware decoding or multi-turn interaction mechanisms that allow the system to request clarification when such ambiguities arise.

\textbf{Incorrect cell values.}  We observed that in 10\% of failures, the predicted SQL queries referenced incorrect cell values due to exact string matching, which often resulted in an empty table. Typical examples include mismatches in label specificity or casing, such as ``height'' versus ``height (cm)'' for patient height, and ``M'' versus ``m'' for male gender. As with incorrect ranking errors, these failures highlight the model's difficulty in handling uncertainty and subtle mismatches between natural language and database-specific values. Incorporating mechanisms such as fuzzy string matching or value normalization could help improve robustness and reliability.

\textbf{Incorrect dealing with multiplicity.} Another observed failure mode concerns the handling of multiplicity in query answers. In some cases, the predicted SQL used a subquery that returned multiple rows in a context that expected a scalar value, which results in an invalid query. In other cases, the NL question permits multiple valid rows in the result table, but the predicted query incorrectly restricts the output to a single row.

\textbf{Abstained or truncated output.} The system also abstained from answering or produced truncated outputs for answerable questions, accounting for 6\% of the failure cases.

\textbf{Miscellaneous.} Other less frequent error modes include incorrect output formats, incorrect table or column names, and arithmetic or aggregation errors.

Overall, our error analysis shows that failure cases in CELEC Text-to-SQL generation exhibit consistent and interpretable patterns. These patterns reveal limitations in the model's ability to adapt to emergent SQL dialects and to handle ambiguity in NL queries, suggesting directions for future work.

\section{Discussion \& Conclusion}
\label{sec:conclusion}

We introduced CELEC, a large language model–powered privacy-constrained framework for automated EHR data extraction and analytics. CELEC translates natural language questions into executable SQL queries and can generate simple visualizations of query results, lowering the expertise required to access and explore EHR data. Importantly, CELEC adheres to strict privacy protocols: the LLM accesses only database metadata and does not access patient-level data. In an evaluation on answerable MIMIC-IV cases in the EHRSQL-2024 benchmark, CELEC achieved an RS(0) score of 81.05\%, comparable to existing systems, while adhering to a privacy-conscious design and maintaining low latency and low cost. Our ablation studies further highlighted the critical role of few-shot demonstrations, schema grounding, reasoning-optimized base models, and lightweight retry mechanisms in achieving robust SQL generation. Together, these results demonstrate that CELEC is an effective and practical approach in curating easily accessible EHR datasets. It has high potential to bridge the accessibility gap between researchers and large-scale EHR databases and to streamline workflows and accelerate discovery in healthcare research.

Several limitations of our work suggest directions for future research. First, our evaluation has limitations in both benchmarking and component-level assessment. To ensure executability, we filtered out unanswerable EHRSQL queries, which prevents direct comparison with systems evaluated on the original benchmark splits; additionally, we do not evaluate the visualization module and provide quantitative metrics to assess the correctness or appropriateness of the generated visualizations. Developing standardized evaluation criteria for data visualization in text-to-SQL pipelines is an important direction for future work. Second, 
our system is evaluated exclusively on the MIMIC database schema. While our system is not architecturally tied to MIMIC-specific table structures and terminology, further validation on additional EHR databases would help assess CELEC's robustness and portability across different settings. Finally, we do not conduct real-world deployment evaluations. While the system is designed to lower technical barriers for clinical researchers, future user studies would help determine its practical utility, usability, and impact on research workflows.

	\bibliographystyle{plain}
	\bibliography{reference}
	
	\newpage
	\appendix
\section{Prompt for SQL Generation}
\label{apd:prompt-template-sql}

Listing~\ref{lst:sql-generation} shows the prompt template we use to generate SQL code from an NL question. The placeholders {\tt \{table\_info\}}, {\tt \{fewshot\_demo\}}, and {\tt \{question\}} correspond to schema information, few-shot demonstrations, and the input NL question, respectively. A SQL query wrapper guides the LLM to produce structured output, enabling post-processing components to reliably extract the generated SQL query using regular expression matching.

Listing~\ref{lst:schema-information} shows the template used to format schema information. After the table name, the schema is presented in tabular form, where each row represents an attribute in the database. The columns {\tt name}, {\tt type}, {\tt notnull}, {\tt dflt\_value}, and {\tt pk} indicate the column name, declared data type, whether the column has a non-null constraint, its default value, and whether it is a primary key. Each table is formatted in this way, and tables are listed sequentially in lexicographical order.

Listing~\ref{lst:few-shot-demo} shows the template for formatting few-shot demonstrations. After the demo NL question, we include a chain-of-thought step beginning with the phrase ``Let's think step-by-step,'' followed by a list of relevant tables referenced in the demo SQL query. The demo SQL query itself is then presented. All few-shot demos in a prompt are displayed in this format and ordered by decreasing similarity of question embeddings, as described in the main text.

\begin{lstlisting}[caption={Prompt template for SQL generation.},
	label={lst:sql-generation},
	float=hbtp, breaklines=true, breakatwhitespace=true,
	basicstyle=\ttfamily\small, frame=single]
	### You are a DuckDB expert.
	Given an input question, write a syntactically correct DuckDB query that answers the input question.
	You must write the query using only functions and syntax supported by DuckDB.
	Even though some examples below (the few-shot demos) include table names prefixed with "physionet-data." and may use functions from other SQL dialects, your final query must:
	- Use only DuckDB-compatible functions (for example, use DATEDIFF instead of TIMESTAMP_DIFF).
	- Use local table names that are capitalized and do NOT include the "physionet-data." prefix.
	- Never query for all columns from a table - only query the columns needed to answer the question.
	- Wrap each column name in backticks (`) to denote them as delimited identifiers.
	- Only reference column names that exist in the tables provided below.
	- Be careful not to query for columns that do not exist, and ensure you are using the correct columns for each table.
	- Use CURRENT_DATE() if the question involves "today".
	- Exclude null values.
	- Use aggregation functions like COUNT() and a GROUP BY clause if the question asks for distribution.
	- Use different aliases for all tables and output columns.
	Wrap the SQL query like this:
	```sql
	```
	
	### Here is the information about the tables:
	{schema_info}
	
	### Some example pairs of questions and corresponding SQL queries (these examples might use functions from other dialects) are provided below:
	{fewshot_demo}
	
	---
	
	### Question: {question}
	### SQL: Let's think step-by-step.
\end{lstlisting}

\begin{lstlisting}[caption={Schema information for SQL generation for example table.},
	label={lst:schema-information},
	float=hbtp, breaklines=true, breakatwhitespace=true,
	basicstyle=\ttfamily\small, frame=single]
	Table: admissions
	Schema:
	cid                name       type  notnull dflt_value     pk
	0     0              row_id     BIGINT    False       None  False
	1     1          subject_id     BIGINT    False       None  False
	2     2             hadm_id     BIGINT    False       None  False
	3     3           admittime  TIMESTAMP    False       None  False
	4     4           dischtime  TIMESTAMP    False       None  False
	5     5      admission_type    VARCHAR    False       None  False
	6     6  admission_location    VARCHAR    False       None  False
	7     7  discharge_location    VARCHAR    False       None  False
	8     8           insurance    VARCHAR    False       None  False
	9     9            language    VARCHAR    False       None  False
	10   10      marital_status    VARCHAR    False       None  False
	11   11                 age     BIGINT    False       None  False
	
	Table: chartevents
	Schema:
	cid        name       type  notnull dflt_value     pk
	0    0      row_id     BIGINT    False       None  False
	...
	7    7    valueuom    VARCHAR    False       None  False
	
	Table: cost
	...
	...
\end{lstlisting}

\begin{lstlisting}[caption={Example few-shot demo for SQL generation.},
	label={lst:few-shot-demo},
	float=hbtp, breaklines=true, breakatwhitespace=true,
	basicstyle=\ttfamily\small, frame=single]
	## 
	Question: What is the minimum total hospital cost that involved a procedure called other enterostomy since 2100?
	Answer: Let's think step-by-step.
	1. Identify the relevant tables:
	-- cost
	-- procedures_icd
	-- d_icd_procedures
	2. Final SQL query:
	SELECT MIN(T1.C1) FROM (SELECT SUM(cost.cost) AS C1 FROM cost WHERE cost.hadm_id IN (SELECT procedures_icd.hadm_id FROM procedures_icd WHERE procedures_icd.icd_code = (SELECT d_icd_procedures.icd_code FROM d_icd_procedures WHERE d_icd_procedures.long_title = 'other enterostomy')) AND STRFTIME(CAST(cost.chargetime AS TIMESTAMP), '%Y') >= '2100' GROUP BY cost.hadm_id) AS T1
	
	## 
	Question: What is the maximum total hospital cost associated with postprocedural pneumothorax in 2100?
	Answer: Let's think step-by-step.
	1. Identify the relevant tables:
	-- cost
	-- diagnoses_icd
	-- d_icd_diagnoses
	2. Final SQL query:
	SELECT MAX(T1.C1) FROM (SELECT SUM(cost.cost) AS C1 FROM cost WHERE cost.hadm_id IN (SELECT diagnoses_icd.hadm_id FROM diagnoses_icd WHERE diagnoses_icd.icd_code = (SELECT d_icd_diagnoses.icd_code FROM d_icd_diagnoses WHERE d_icd_diagnoses.long_title = 'postprocedural pneumothorax')) AND STRFTIME(CAST(cost.chargetime AS TIMESTAMP), '%Y') = '2100' GROUP BY cost.hadm_id) AS T1
\end{lstlisting}

\section{Prompt for Visualization Generation}
\label{apd:prompt-template-viz}

Listing~\ref{lst:viz-generation} shows the prompt template we use to generate a visualization from an NL question and the corresponding extracted dataframe. The placeholders {\tt \{viz\_name\}}, {\tt \{columns\}}, and {\tt \{question\}} correspond to (1) the list of visualization types supported by our system's front end (scatterplot, bar chart, line chart, and histogram), (2) the columns of the extracted dataframe, and (3) the input NL question, respectively.

\begin{lstlisting}[caption={Prompt template for visualization generation},
	label={lst:viz-generation},
	float=hbtp, breaklines=true, breakatwhitespace=true,
	basicstyle=\ttfamily\small, frame=single]
	### You are an expert in data visualization. Given an input question and the columns of a dataframe, determine the type of visualization and the axis columns that most appropriately address the question.
	For the type of visualization, select only among the following options: {viz_names}. For the axis columns, select only the column names you are given.
	Select only one column if the visualization type requires only one column of data (e.g., histogram), and select two columns if the visualization type requires two columns of data (e.g., scatterplot)
	If the prompt is related to visualizing a distribution and there is a count column, choose the bar chart and select two columns instead of selecting the histogram or density plot and only one column.
	
	Format the output like the following without any extra text or explanations (the Yaxis segment can be omitted if only one column is selected): 
	VizType: VISUALIZATION-TYPE-NO; Xaxis: HORIZONTAL-AXIS; Yaxis: VERTICAL-AXIS
	As an example, the output "VizType: 0; Xaxis: cal_daily; Yaxis: bmi" means a scatterplot with the column "cal_daily" as the horizontal axis and the column "bmi" as the vertical axis.
	
	### Here are the column names in the given dataframe.
	{columns}
	
	### Question: {question}
	### Answer:
\end{lstlisting}

\section{Medical Literature–Inspired Demos}
\label{apd:medical-lit-demo}

To supplement benchmark data, we created 105 few-shot demonstrations inspired by published studies that used the MIMIC-IV dataset. Candidate cohort criteria were identified by searching for ``MIMIC-IV'' on PubMed and reviewing approximately 20 recent papers. From these, we adapted subject selection descriptions into natural language questions. For example, a criterion such as ``We selected patients that…'' was paraphrased into ``Select patients that…,'' while visualization-oriented questions were framed in formats like ``Generate the distribution of … for all patients that….'' In addition to direct adaptations, some questions were modified to expand coverage across different database attributes or application scenarios (e.g., visualization as well as cohort selection).

The corresponding gold SQL queries were first generated by {\tt gpt-4-1106-preview} and then validated against DuckDB implementations of the MIMIC databases. We manually verified their correctness in our local database system, defined as valid, free of execution errors, and accurately presenting the information needed to address the corresponding question. Most of the debugging process consisted of resolving execution errors, but also included more arduous fixes, such as rewriting the entire query.

All demos follow the same format as described in Appendix~\ref{apd:prompt-template-sql}, including chain-of-thought reasoning steps and explicit table selection, ensuring consistency across sources. This process yielded 105 high-quality demonstrations that complement benchmark-derived examples and better reflect clinically realistic query patterns.

\section{EHRSQL Data Point Preprocessing}
\label{apd:test-set-sampling}

We modified the train, validation, and test sets of EHRSQL, particularly the gold SQL queries, to ensure compatibility with CELEC's design and execution environment. Since the original EHRSQL benchmark runs on a SQLite version of the MIMIC-IV demo database, while CELEC operates on DuckDB instances of MIMIC-III and MIMIC-IV, we adapted the data through a combination of automated translation and manual correction. We also filtered out unanswerable questions to focus on executable, clinically relevant queries. Below, we describe these preprocessing steps.

\subsection{Translating between SQL dialects}

We first applied the SQLGlot package in Python to translate gold SQL queries from SQLite into DuckDB syntax. While SQLGlot handled many standard cases automatically, it failed to account for several recurring issues. To improve compatibility, we introduced a set of targeted preprocessing rules that addressed the most frequently observed errors, as summarized below:

\begin{itemize}
	\item \textbf{Current time.}  
	SQLite uses the keyword \texttt{CURRENT\_TIME} to return the current time of day, whereas DuckDB expects \texttt{CURRENT\_TIMESTAMP}, which returns both date and time. SQLGlot left these untranslated, so we replaced all instances of \texttt{CURRENT\_TIME} with \texttt{CURRENT\_TIMESTAMP}.
	
	\item \textbf{Datetime expressions.}  
	SQLite allows flexible modifiers within \texttt{datetime()} calls, such as ``start of month'' or ``+1 day.'' SQLGlot did not handle these consistently, so we introduced explicit mappings:  
	\begin{itemize}
		\item \texttt{datetime(expr)} → \texttt{CAST(expr AS TIMESTAMP)}  
		\item \texttt{datetime(expr, 'start of X')} → \texttt{date\_trunc('X', expr)}  
		\item \texttt{datetime(expr, '+/-N unit')} → \texttt{expr +/- INTERVAL 'N unit'}  
		\item \texttt{datetime(expr, '-0 year')} → no-op (left unchanged)  
	\end{itemize}
\end{itemize}

These issues were not isolated: each of the error types above appeared more than 500 times in the dataset. By handling them systematically, we corrected the vast majority of translation failures. Regarding other translational errors, due to their rarity and heterogeneity, we did not attempt case-level debugging and discarded them in the subsequent filtering stage.

\subsection{Discarding unanswerable questions}

After translation, we ensured that all splits contained only answerable queries, since detecting unanswerable ones is not a primary goal of CELEC. Unanswerable cases arose from two sources. First, the original EHRSQL benchmark deliberately included unanswerable questions as a separate evaluation task. Second, some queries became unanswerable under our setup: although answerable on the SQLite-based benchmark database, they failed on the official MIMIC-IV demo database when executed in DuckDB. Such failures included gold queries that produced execution errors or queries that executed but returned an empty dataframe. Our preprocessing identified 609 queries that consistently yielded empty dataframes; these were excluded from further use. 

We discarded all such cases across the training, validation, and test sets, retaining only those queries where the gold SQL executed successfully on the official MIMIC-IV demo database and returned a non-empty dataframe. This filtering process resulted in new dataset splits comprising 3,976 training queries, 785 validation queries, and 785 test queries. For system development, we further divided the new test set into a small validation portion of 78 queries (10\%) for hyperparameter tuning and 707 queries (90\%) for final evaluation, as described in the main text. 

This two-step preprocessing pipeline, i.e., translation followed by filtering, ensured that the datasets used in CELEC are coherent with DuckDB's execution environment and focus exclusively on answerable, clinically meaningful queries.

\section{User Interface Demonstration}
\label{apd:ui-demo}
\begin{figure}[hbtp]
	\vspace{-10pt}
	\centering
	\includegraphics[width=0.7\textwidth]{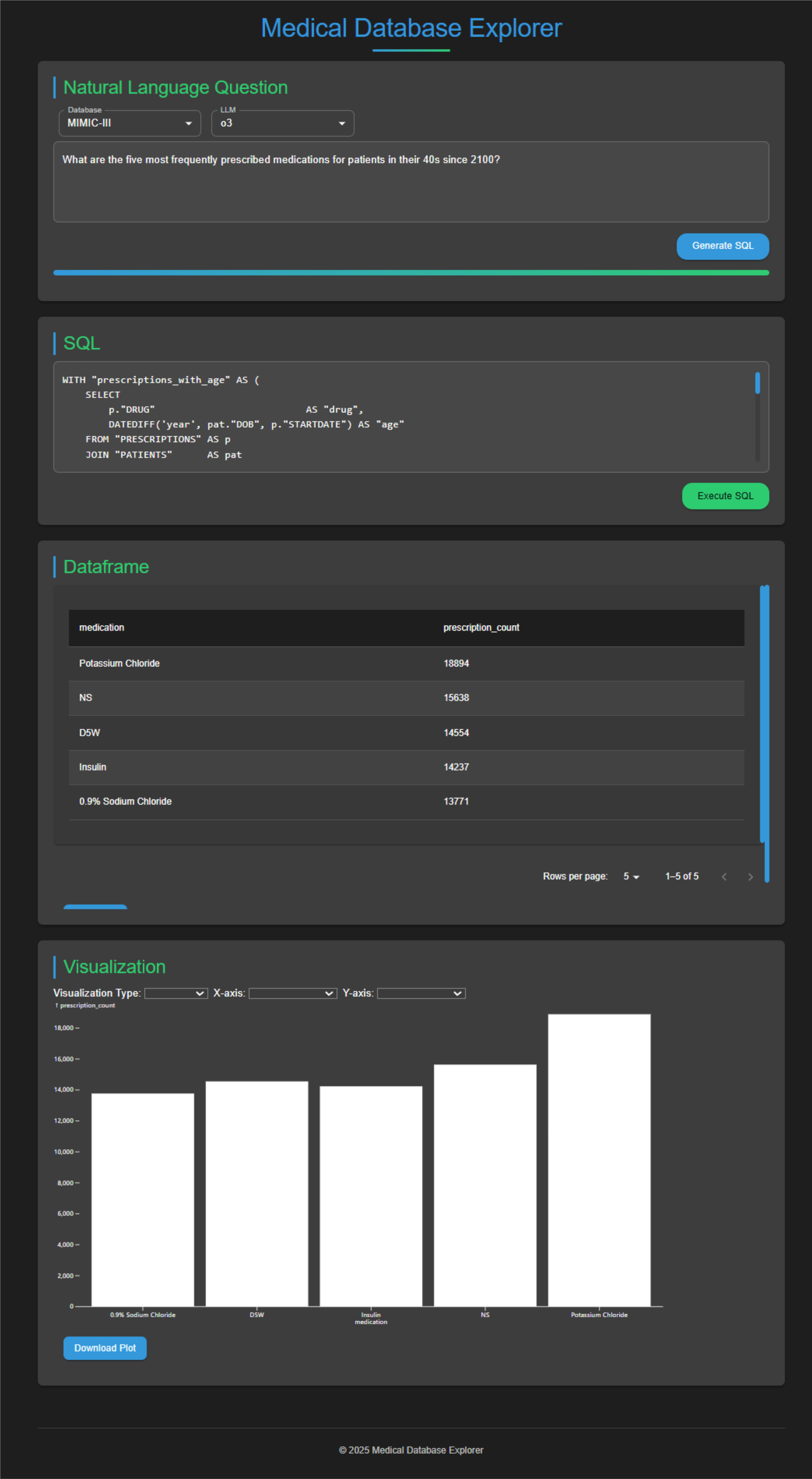}
	\caption{CELEC user interface, with results of running on an example question}
	\vspace{-10pt}
	\label{fig:ui}
\end{figure}

Figure \ref{fig:ui} illustrates the CELEC user interface, with results of running on the example question ``What are the five most frequently prescribed medications for patients in their 40s since 2100?'' The system enables users to formulate database queries by entering a natural language question and selecting the appropriate database and LLM backbone. Upon submission, the corresponding SQL query is automatically generated. The interface further provides modules for inspecting the resulting dataframe, visualizing query outputs, and constructing cohort selection flowcharts.

\end{sloppypar}

\end{document}